\documentstyle[amsfonts,12pt,sw20lart]{article}

\input tcilatex
\QQQ{Language}{
American English
}

\begin{document}

\author{Nikolai Laskin\thanks{%
E-mail: nlaskin@rocketmail.com}}
\title{{\bf Fractional Quantum Mechanics and L\'evy Path Integrals}}
\date{Isotrace Laboratory, University of Toronto\\
60 St. George Street, Toronto, ON M5S 1A7\\
Canada }
\maketitle

\begin{abstract}
The fractional quantum and statistical mechanics have been developed via new
path integrals approach.

{\it PACS }number(s): 03.65.-w, 05.30.-d, 05.40. Fb, 03.65. Db
\end{abstract}

From a functional integration point of view the quantum mechanics in
Feynman's path integrals treatment \cite{Feynman}, \cite{Grosche} is the
theory of functionals on measure generated by the Brownian motion-Wiener
stochastic process \cite{Wiener}, \cite{Kac}. The Wiener process is the
Markov, Gaussian or normal stochastic process the statistical description of
which is given by well-known diffusion equation \cite{Kac}.

The natural generalization of the Brownian motion is the L\'evy flights (or
L\'evy motion). The mathematical foundation of this generalization is the
theory of stable probability distributions developed by L\'evy \cite{Levy}.
The most fundamental property of the L\'evy distributions is the stability
under addition, following from the generalized central limit theorem valid
for L\'evy distributions. So, from the probability theory point of view the
stable probability law is the generalization of well-known Gaussian law.

L\'evy flights is widely used to model a variety processes such as anomalous
diffusion \cite{Mandelbrot}, turbulence \cite{Klafter}, chaotic dynamics 
\cite{Zaslavsky}, plasma physics \cite{Zimbardo}, financial dynamics \cite
{Mantega}, biology and physiology\cite{West}. The most recent studies are
related to fractional L\'evy motion \cite{Chechkin} which is the
generalization of the fractional Brownian motion \cite{Mandelbrot}.

This paper deals with new path integrals over measures induced by the
stochastic process of L\'evy flights. I have generalized the Feynman and
Wiener path integrals and developed the new fractional quantum mechanics
(fQM) and fractional statistical mechanics (fSM). The fQM and fSM include
the Feynman quantum and statistical mechanics in similar way as the L\'evy
process (L\'evy flights) generalizes the Wiener process (Brownian motion).

In this paper I have developed the foundations of fQM and fSM in their
functional formulations. As the next step we should think about the concrete
fQM problems such as, for example, fractional harmonic oscillator,
fractional hydrogen atom, fractional scattering theory, relativistic
fractional quantum mechanics, fractional quantum gauge field theory etc.

{\bf 1.} The Feynman quantum mechanics deals with Feynman's quantum
mechanical functionals $K_F(x_bt_b|x_at_a)$ \cite{Feynman}

\begin{equation}
\stackrel{\cdot }{K_F(x_bt_b|x_at_a)=\int\limits_{x(t_a)=x_a}^{x(t_b)=x_b}%
{\cal D}_{Feynman}x(\tau )\cdot \exp \{\frac i\hbar
\int\limits_{t_a}^{t_b}d\tau V(x(\tau ))\}},  \label{eq1}
\end{equation}

where $V(x(\tau ))$ is the potential energy as a functional of a particle
path $x(\tau )$ and the Feynman functional measure is defined as

\begin{equation}
\int\limits_{x(t_a)=x_a}^{x(t_b)=x_b}{\cal D}_{Feynman}x(\tau )....=%
\stackunder{N\rightarrow \infty }{\lim }\int dx_1...dx_{N-1}\left( \frac{%
2\pi i\hbar \varepsilon }m\right) ^{-N/2}\times  \label{eq2}
\end{equation}

\[
\times \prod\limits_{j=1}^N\exp \left\{ \frac{im}{2\hbar \varepsilon }%
(x_j-x_{j-1})^2\right\} ..., 
\]

here $m$ is the mass of the quantum mechanical particle, $\hbar $ is the
Planck's constant, $x_0=x_a$, $x_N=x_b$ and $\varepsilon =(t_b-t_a)/N$.

I propose the fractional quantum mechanics based on the new path integral

\begin{equation}
\stackrel{\cdot }{K_L(x_bt_b|x_at_a)=\int\limits_{x(t_a)=x_a}^{x(t_b)=x_b}%
{\cal D}_{Laskin}x(\tau )\cdot \exp \{\frac i\hbar
\int\limits_{t_a}^{t_b}d\tau V(x(\tau ))\}},  \label{eq3}
\end{equation}

where $V(x(\tau ))$ is the potential energy as a functional of the L\'evy
particle path and the Laskin path integral measure is defined as

\begin{equation}
\int\limits_{x(t_a)=x_a}^{x(t_b)=x_b}{\cal D}_{Laskin}x(\tau )...=
\label{eq4}
\end{equation}

\[
=\stackunder{N\rightarrow \infty }{\lim }\int dx_1...dx_{N-1}\left( \frac{%
iD_\alpha \varepsilon }\hbar \right) ^{-N/\alpha }\cdot
\prod\limits_{j=1}^NL_\alpha \left\{ \left( \frac \hbar {iD_\alpha
\varepsilon }\right) ^{1/\alpha }|x_j-x_{j-1}|\right\} ..., 
\]

where $D_\alpha $ is the generalized ''diffusion coefficient'', $\hbar $
denotes the Planck's constant, $x_0=x_a$, $x_N=x_b$, $\varepsilon
=(t_b-t_a)/N$, and the L\'evy function $L_\alpha $ is expressed in terms of
Fox's $H$ functions\cite{Fox}, \cite{Mathai}, \cite{West1}

\begin{equation}
(\frac{D_\alpha t}\hbar )^{-1/\alpha }L_\alpha \left\{ \left( \frac \hbar
{D_\alpha t}\right) ^{1/\alpha }|x|\right\} =  \label{eq5}
\end{equation}

\[
=\frac \pi {\alpha |x|}H_{2,2}^{1,1}\left[ \left( \frac \hbar {D_\alpha
t}\right) ^{1/\alpha }|x|\mid \QATOP{(1,1/\alpha ),(1,1/2)}{(1,1),(1,1/2)}%
\right] , 
\]

here $\alpha $ is the L\'evy index, $0<\alpha \leq 2$.

The Eqs (\ref{eq3})-(\ref{eq5}) define the fractional quantum mechanics via
Laskin path integral.

The fractional free particle propagator $K_L^{(0)}(x_bt_b|x_at_a)$ has the
form

\begin{equation}
K_L^{(0)}(x_bt_b|x_at_a)=\int\limits_{x(t_a)=x_a}^{x(t_b)=x_b}{\cal D}%
_{Laskin}x(\tau )\cdot 1=  \label{eq6}
\end{equation}

\[
=\left( \frac{iD_\alpha (t_b-t_a)}\hbar \right) ^{-1/\alpha }L_\alpha
\left\{ \left( \frac \hbar {iD_\alpha (t_b-t_a)}\right) ^{1/\alpha
}|x_b-x_a|\right\} . 
\]

It is known that at $\alpha =2$ the L\'evy distribution is transformed to
Gaussian and the L\'evy flights process is transformed to the process of the
Brownian motion. The Eq.(\ref{eq6}) in accordance with definition given by
the Eq.(\ref{eq5}) and the properties of the Fox's function $H_{2,2}^{1,1}$
at $\alpha =2$ (see, \cite{Mathai}, \cite{West1}) is transformed to Feynman
free particle propagator\cite{Feynman}

\begin{equation}
K_F^{(0)}(x_bt_b|x_at_a)=\left( \frac{2\pi i\hbar (t_b-t_a)}m\right)
^{-1/2}\cdot \exp \left\{ \frac{im(x_b-x_a)^2}{2\hbar (t_b-t_a)}\right\} .
\label{eq7}
\end{equation}

In terms of Fourier integral (momentum representation) the propagator $%
K_L^{(0)}(x_bt_b|x_at_a)$ is written as follows

\begin{equation}
K_L^{(0)}(x_bt_b|x_at_a)=\frac 1{2\pi \hbar }\int dp\cdot \exp \left\{ i%
\frac{p(x_b-x_a)}\hbar -i\frac{D_\alpha (t_b-t_a)|p|^\alpha }\hbar \right\} .
\label{eq8}
\end{equation}

We see that the energy $E_p$ of the fractional quantum mechanical particle
with the momentum $p$ is given by

\begin{equation}
E_p=D_\alpha |p|^\alpha .  \label{eq9}
\end{equation}

The comparison of the Eq.(\ref{eq7}) and the Eq.(\ref{eq8}) allows to
conclude that at $\alpha =2$ we should put $D_2=1/2m$. Then the Eq.(\ref{eq9}%
) is transformed to $E_p=p^2/2m$.

Using the Eq.(\ref{eq9}) we can define the {\it Laskin functional measure in
the momentum representation} by the following way

\begin{equation}
\int\limits_{x(t_a)=x_a}^{x(t_b)=x_b}{\cal D}_{Laskin}x(\tau )...=
\label{eq10}
\end{equation}

\[
=\stackunder{N\rightarrow \infty }{\lim }\int dx_1...dx_{N-1}\frac 1{(2\pi
\hbar )^N}\int dp_1...dp_N\cdot \exp \left\{ i\frac{p_1(x_1-x_a)}\hbar -i%
\frac{D_\alpha \varepsilon |p_1|^\alpha }\hbar \right\} \times ... 
\]

\[
\times \exp \left\{ i\frac{p_N(x_b-x_{N-1})}\hbar -i\frac{D_\alpha
\varepsilon |p_N|^\alpha }\hbar \right\} .... 
\]

here $\varepsilon =(t_b-t_a)/N$.

Taking into account the Eq.(\ref{eq8}) it is easy to check on the
consistency condition

\begin{equation}
K_L^{(0)}(x_bt_b|x_at_a)=\int dx^{\prime }K_L^{(0)}(x_bt_b|x^{\prime
}t^{\prime })\cdot K_L^{(0)}(x^{\prime }t^{\prime }|x_at_a).  \label{eq11}
\end{equation}

The fractional free particle propagator is governed by the fractional
Schr\"odinger equation

\begin{equation}
i\hbar \frac \partial {\partial t_b}K_L^{(0)}(x_bt_b|x_at_a)=-D_\alpha
(\hbar \nabla _b)^\alpha K_L^{(0)}(x_bt_b|x_at_a),  \label{eq12}
\end{equation}

where the Riesz quantum fractional derivative ($\hbar \nabla )^\alpha $ is
defined through its Fourier transform (see, for example, \cite{Oldham})

\begin{equation}
(\hbar \nabla )^\alpha ...=-\frac 1{2\pi \hbar }\int dpe^{i\frac{px}\hbar
}|p|^\alpha ....  \label{eq13}
\end{equation}

The propagator $K_L(x_bt_b|x_at_a)$ describes the evolution of the
fractional quantum-mechanical system

\begin{equation}
\psi _f(x_b,t_b)=\int dx_aK_L(x_bt_b|x_at_a)\cdot \psi _i(x_a,t_a),
\label{eq14}
\end{equation}

where $\psi _i(x_a,t_a)$ is the fractional wave function of initial (at the $%
t=t_a)$ state and $\psi _f(x_b,t_b)$ is the fractional wave function of
final (at the $t=t_b)$ state.

Taking into account the Eqs.(\ref{eq3}), (\ref{eq14}) and the definition (%
\ref{eq13}) we can derive the fractional Schr\"odinger equation for the wave
function of quantum mechanical system

\begin{equation}
i\hbar \frac{\partial \psi }{\partial t}=-D_\alpha (\hbar \nabla )^\alpha
\psi +V(x)\psi .  \label{eq15}
\end{equation}
It is easy to see that the fractional (or Levy) free particle wave function
has the form

\begin{equation}
\psi (x,t)=\frac 1{2\pi \hbar }\int dp\cdot \exp \left\{ i\frac{px}\hbar -i%
\frac{D_\alpha t|p|^\alpha }\hbar \right\} .  \label{eq16}
\end{equation}

Thus, the Eqs.(\ref{eq3}), (\ref{eq4}) and (\ref{eq14}), (\ref{eq15}) are
the basic equations of new fQM. They should be analyzed for the particular
fractional quantum systems.

{\bf 2.} In order to develop the fractional statistical mechanics let us go
in the previous consideration from imaginary time to ''inverse temperature'' 
$\beta =1/k_BT,$ where $k_B$ is the Boltzmann's constant and $T$ is the
temperature, $it\rightarrow \beta $. In the framework of traditional
functional approach to the statistical mechanics we deal with functionals
over the Wiener measure\cite{Feynman}, \cite{Kac}, \cite{Feynman1}

\begin{equation}
\rho _W(x,\beta |x_0)=\int\limits_{x(0)=x_0}^{x(\beta )=x}{\cal D}%
_{Wiener}x(u)\cdot \exp \{-\int\limits_0^\beta duV(x(u))\},  \label{eq17}
\end{equation}

where $\rho _W(x,\beta |x_0)$ is the density matrix of the statistical
system in the external field $V(x)$ and the Wiener measure is given by

\begin{equation}
\int\limits_{x(0)=x_0}^{x(\beta )=x}{\cal D}_{Wiener}x(u)...=\stackunder{%
N\rightarrow \infty }{\lim }\int dx_1...dx_{N-1}\left( \frac{2\pi \hbar
^2\varsigma }m\right) ^{-N/2}\times   \label{eq18}
\end{equation}

\[
\times \prod\limits_{j=1}^N\exp \left\{ -\frac m{2\hbar ^2\varsigma
}(x_j-x_{j-1})^2\right\} ...,
\]

here $\varsigma =\beta /N$ and $x_N=x$.

The fSM deals with L\'evy or fractional density matrix $\rho _L(x,\beta |x_0)
$ which is given by

\begin{equation}
\rho _L(x,\beta |x_0)=\int\limits_{x(0)=x_0}^{x(\beta )=x}{\cal D}_{L\acute
evy}x(u)\cdot \exp \{-\int\limits_0^\beta duV(x(u))\},  \label{eq19}
\end{equation}

where the L\'evy functional measure is defined as follows

\begin{equation}
\int\limits_{x(0)=x_0}^{x(\beta )=x}{\cal D}_{L\acute evy}x(u)...=%
\stackunder{N\rightarrow \infty }{\lim }\int dx_1...dx_{N-1}(D_\alpha
\varsigma )^{-N/\alpha }\times   \label{eq20}
\end{equation}

\[
\prod\limits_{j=1}^NL_\alpha \left\{ \frac{|x_j-x_{j-1}|}{(D_\alpha
\varsigma )^{1/\alpha }}\right\} ...,
\]

here $\varsigma =\beta /N$, $x_N=x$ and the L\'evy function $L_\alpha $ is
given by the Eq.(\ref{eq5}).

It was shown by L\'evy \cite{Levy1} (and later by Doob \cite{Doob}) that the
Eq.(\ref{eq20}) defines the functional measure in the space ${\frak D}$ of
left (or right) continuous functions having only discontinuities of the
first kind.

In Fourier space the fractional density matrix $\rho _L^{(0)}(x,\beta |x_0)$
of a free particle ($V=0$) has the form

\begin{equation}
\rho _L^{(0)}(x,\beta |x_0)=\frac 1{2\pi \hbar }\int dp\exp \left\{ i\frac{%
p(x-x_0)}\hbar -\beta D_\alpha |p|^\alpha \right\} .  \label{eq21}
\end{equation}

When $\alpha =2$ and $D_2=1/2m$ the Eq.(\ref{eq21}) gives the well-known
density matrix for one-dimensional free particle \cite{Feynman1}

\begin{equation}
\rho _W^{(0)}(x,\beta |x_0)=\left( \frac m{2\pi \hbar ^2\beta }\right)
^{1/2}\exp \left\{ -\frac m{2\hbar ^2\beta }(x-x_0)^2\right\} .  \label{eq22}
\end{equation}

The density matrix $\rho _L(x,\beta |x_0)$ obeys the fractional differential
equation

\begin{equation}
-\frac{\partial \rho _L(x,\beta |x_0)}{\partial \beta }=-D_\alpha (\hbar
\nabla _x)^\alpha \rho _L(x,\beta |x_0)+V(x)\rho _L(x,\beta |x_0).
\label{eq23}
\end{equation}

The Eqs. (\ref{eq19}) - (\ref{eq21}) and (\ref{eq23}) are the basic
equations of new fSM in its functional formulation.

{\bf 3.} I have developed the new path integral approach to fQM and fSM. My
approach is based on the new functional measures generated by the stochastic
process of the L\'evy flights. The {\it Laskin} and {\it L\'evy} path
integrals are the generalization of the {\it Feynman} and {\it Wiener} path
integrals respectively.

It will be interesting to find the physical systems and objects the quantum
dynamics (statistical mechanics) of which is governed by the developed fQM
(fSM).

\end{document}